\documentclass[a4paper]{article}
\pdfoutput=1
\usepackage{INTERSPEECH2021}
\usepackage{xspace}
\usepackage{tikz}
\usepackage{algorithm,algorithmicx,algpseudocode}
\usepackage{mathtools}
\usepackage{amsmath, amssymb}
\usetikzlibrary{calc, shapes}
\usetikzlibrary{decorations.text}
\usetikzlibrary{arrows.meta}
\usepackage{hyperref}
\usepackage{url}
\usepackage{multirow}
\usepackage{tabularx}
\usepackage{subcaption}
\usepackage{siunitx,mwe}

\title{NU-Wave: A Diffusion Probabilistic Model for Neural Audio Upsampling}
\name{Junhyeok Lee$^1$, Seungu Han$^{1, 2}$}
\address{$^1$MINDs Lab Inc., Republic of Korea \qquad $^2$Seoul National University,  Republic of Korea}
\email{\{jun3518, hansw0326\}@mindslab.ai}

\makeatletter
\DeclareRobustCommand\onedot{\futurelet\@let@token\@onedot}
\def\@onedot{\ifx\@let@token.\else.\null\fi\xspace}

\def\etal{\emph{et al}\onedot}
\makeatother

\newcommand{\inw}{\textit{NU-Wave}\xspace}
\newcommand{\nw}{NU-Wave\xspace}
\newcommand{\asr}{audio super-resolution\xspace}
\begin{document}

\maketitle

\begin{abstract}
In this work, 
we introduce \inw, the first neural audio upsampling model to produce waveforms of sampling rate 48kHz from coarse 16kHz or 24kHz inputs, while prior works could generate only up to 16kHz.
\nw is the first diffusion probabilistic model for \asr which is engineered based on neural vocoders.
\nw generates high-quality audio that achieves high performance in terms of signal-to-noise ratio (SNR), log-spectral distance (LSD), and accuracy of the ABX test.
In all cases, \nw outperforms the baseline models despite the substantially smaller model capacity (3.0M parameters) than baselines (5.4-21\%).
The audio samples of our model are available at \url{https://mindslab-ai.github.io/nuwave}, and the code will be made available soon.

\end{abstract}

\setlength{\textfloatsep}{12pt}%

\section{Introduction}
Audio super-resolution, neural upsampling, or bandwidth extension is the task of generating high sampling rate audio signals with full frequency bandwidth from low sampling rate signals.  
There have been several works that applied \textit{deep neural networks} to audio super-resolution \cite{dnnbwe,audiosuperres,tfasr,asr4sr,bwegan,tfilm,spkppgbwe}. 
Still, prior works used 16kHz as the target frequency, which is not considered a high-resolution.
Since the highest audible frequency of human is 20kHz, high-resolution audio used in multimedia such as movies or musics uses 44.1kHz or 48kHz.

Similar to recent works in the image domain \cite{prsrgan,esrgan,pulse}, \asr works also used deep generative models \cite{asr4sr,bwegan}, mostly focused on generative adversarial networks (GAN) \cite{gan}.
On the other hand, prior works in neural vocoders, which is inherently similar to audio super-resolution due to its local conditional structure, adopted a variety of generative models such as autoregressive models \cite{wavenet,wavernn}, flow-based models \cite{waveglow,waveflow}, and variational autoencoders (VAE) \cite{wavevae}.
Recently, diffusion probabilistic models \cite{dpm, ddpm} were applied to neural vocoders \cite{diffwave,wavegrad}, with remarkably high perceptual quality.
Details of diffusion probabilistic models and their applications are explained in Section \ref{subsection:dpm} and \ref{subsection:cdm}.

In this paper, we introduce \inw, a conditional diffusion probabilistic model for neural audio upsampling.
Our contributions are as follows:
\begin{enumerate}
    \item \nw is the first deep generative neural audio upsampling model to synthesize waveforms of sampling rate 48kHz from coarse 16kHz or 24kHz inputs.
    \item We adopted diffusion probabilistic models for the audio upsampling task for the first time by engineering neural vocoders based on diffusion probabilistic models.
    \item \nw outperforms previous models \cite{audiosuperres,bwegan} including the GAN-based model in both quantitative and qualitative metrics with substantially smaller model capacity (5.4-21\% of baselines).
\end{enumerate}

\section{Related works}
\subsection{Diffusion probabilistic models} \label{subsection:dpm}
Diffusion probabilistic models (diffusion models for brevity) are trending generative models \cite{dpm,ddpm}, which apply iterative Monte-Carlo Markov chain sampling to generate complex data from simple noise distribution such as normal distribution.
Markov chain of a diffusion model consists of two processes: the \textit{forward process}, and the \textit{reverse process}. 

The \textit{forward process}, also referred to as the \textit{diffusion process}, gradually adds Gaussian noise to obtain whitened latent variables $y_1,y_2,\dotsc,y_T \in \mathbb{R}^L$ from the data $y_0 \in \mathbb{R}^L$, where $L$ is time length of the data. Unlike other generative models, such as GAN \cite{gan} and VAE \cite{vae}, the diffusion model's latent variables have the same dimensionality as the input data.
Shol-Dickstein \etal \cite{dpm} applied a fixed noise variance schedule $\beta_{1:T}\coloneqq [ \beta_1,\beta_2,\dotsc,\beta_T ]$ to define the forward process $q(y_{1:T} | y_0)\coloneqq \prod_{t=1}^T q(y_t | y_{t - 1})$ as the multiplication of Gaussian distributions:
\begin{align}
    q(y_t | y_{t - 1}) &\coloneqq \mathcal{N}\left(y_t; \sqrt{1 - \beta_t}\,y_{t - 1}, \beta_t I\right).
\end{align}
Since the sum of Gaussian distributions is also a Gaussian distribution, we can write $y_t$, the latent variable at a timestep $t$, as  $q(y_t | y_{0}) \coloneqq \mathcal{N}\left(y_t; \sqrt{\bar\alpha_t}\,y_{0}, (1-\bar\alpha_t) I\right)$, where $\alpha_0\coloneqq1$, $\alpha_t \coloneqq1-\beta_t$ and $\bar\alpha_t\coloneqq\prod_{s=1}^t \alpha_s$.
$\beta_{1:T}$ are predefined hyperparameters to match the forward and the reverse distributions by minimizing $D_{KL}(q(y_T | y_{0})|| \mathcal{N}(0,I))$.
Ho \etal \cite{ddpm} set these values as a linear schedule of $\beta_{1:T}=\text{Linear}(10^{-4},0.02, 1000)$.

The \textit{reverse process} is defined as the reverse of the diffusion where the model, which is parametrized by $\theta$, learns to estimate the added Gaussian noise.
Starting from $y_T$, which is sampled from Gaussian distribution $p(y_T)=\mathcal{N}\left(y_T;0,I\right)$, the reverse process $p_{\theta}(y_{0:T})\coloneqq p(y_T)\prod_{t=1}^T p_{\theta}(y_{t-1} | y_t)$ is defined as the multiplication of transition probabilities:
\begin{align}
    p_{\theta}(y_{t-1} | y_{t}) &\coloneqq \mathcal{N}\left(y_{t-1}; \mu_\theta(y_t, t) , \sigma^2_t I \right),
\end{align}
where $\mu_{\theta}(y_t,t)$ and $\sigma^2_{t}$ are the model estimated mean and variance of $y_{t-1}$. Ho \etal suggested to set variance as timestep dependant constant.
Latent variable $y_{t-1}$ is sampled from the distribution $p(y_{t-1} | y_t)$ where $\epsilon_\theta$ is model estimated noise and $z\sim \mathcal{N}(0,I)$:
\begin{align}
y_{t-1}=\frac{1}{{\sqrt{\alpha_t}}}{\left( y_t - \frac{1-\alpha_t}{\sqrt{1-\bar\alpha_t}}\, \epsilon_\theta(y_t, t) \right)}+  \sqrt{\frac{1-\bar\alpha_{t-1}}{1-\bar\alpha_t} \beta_t}z.
\end{align}

Diffusion models approximate the probability density of data $p_\theta(y_0)\coloneqq \int p_\theta(y_{0:T})dy_{1:T}$.
The training objective of the diffusion model is minimizing the variational bound on negative log-likelihood without any auxiliary losses.
The variational bound of the diffusion model is represented as KL divergence of the forward process and the reverse process.
Ho \etal \cite{ddpm} reparametrized the variational bound to a simple loss which is connected with denoising score matching and Langevin dynamics \cite{denoisingscorematching,slicedscorematching,improvedscorebased}. For a diffused noise $\epsilon \sim \mathcal{N}(0,I)$, the loss resembles denoising score matching:
\begin{align}
     \mathbb{E}_{t,y_0,\epsilon}\left[\left\lVert \epsilon - \epsilon_\theta\left(\sqrt{\bar \alpha_t}\, y_0 +\sqrt{1 - \bar\alpha_t}\, \epsilon, t\right)\right\rVert_2^2 \right]. \label{eqn:ddpmelbo}
\end{align}

\subsection{Conditional diffusion models as a neural vocoder}\label{subsection:cdm}
Neural vocoders adopted diffusion models \cite{diffwave,wavegrad} by modifying them to a conditional generative model for $x$, where $x$ is a local condition like mel-spectrogram.
The probability density of a conditional diffusion model can be written as
     $p_\theta(y_0 | x)\coloneqq \int p_\theta(y_{0:T}| x)dy_{1:T}$.
Furthermore, Chen \etal \cite{wavegrad} suggested training the model with the continuous noise level $\sqrt{\bar\alpha}$, which is uniformly sampled between adjacent discrete noise levels $\sqrt{\bar\alpha_{t}}$ and  $\sqrt{\bar\alpha_{t-1}}$, instead of the timestep $t$.
Continuous noise level training allows different noise schedules during training and sampling, while a discrete timestep trained model is needed the same schedule.
Along with the continuous noise level condition, Chen \etal \cite{wavegrad} replaced the $L_2$ norm of Eq.\~(\ref{eqn:ddpmelbo}) to $L_1$ norm for empirical training stability.

\section{Approach}

\begin{figure}[t!]
  \centering
  \includegraphics[width=.9\linewidth]{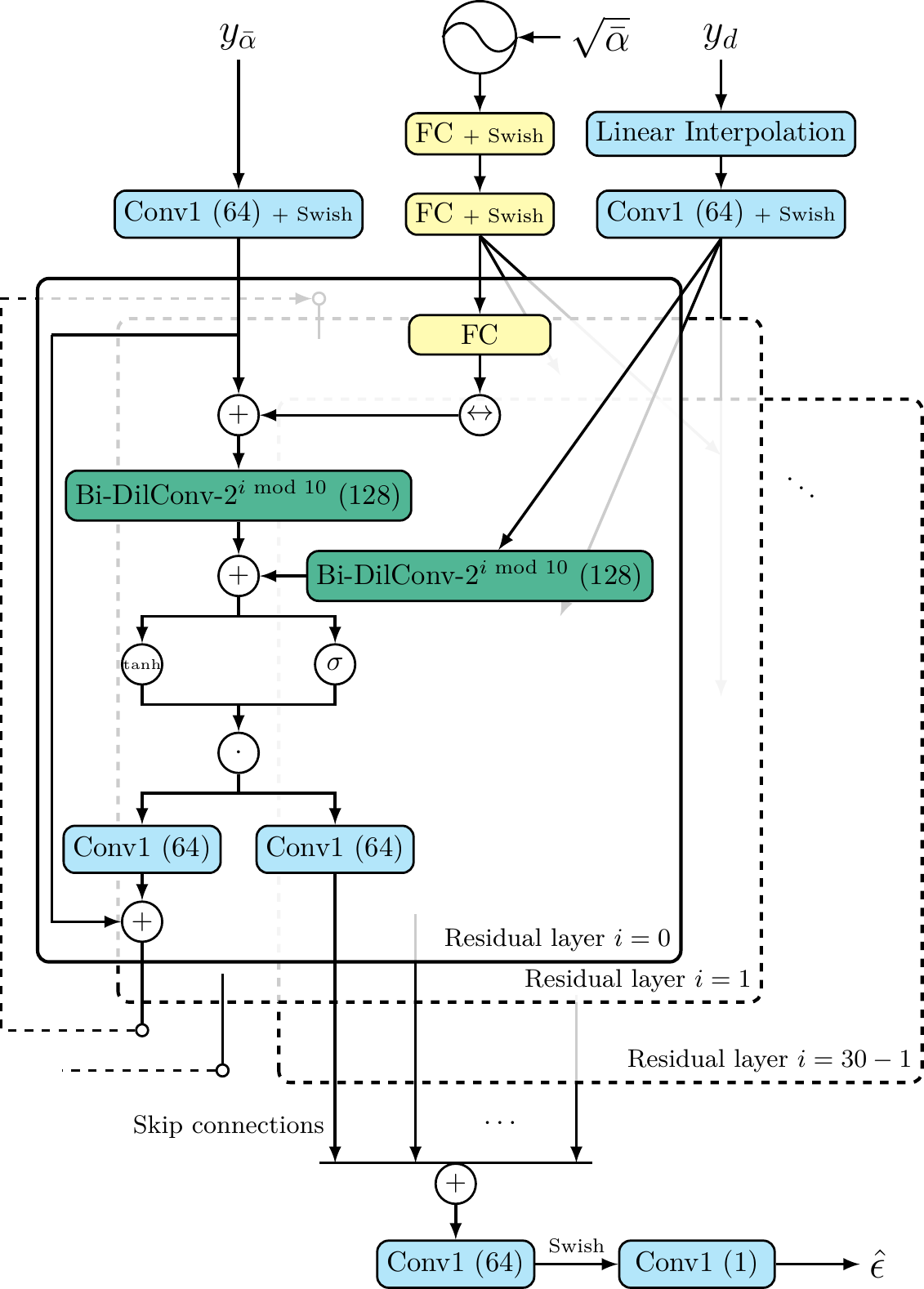}
        \vspace{-0.5\baselineskip}
  \caption{The network architecture of \nw. Noisy speech $y_{\bar\alpha}$, downsampled speech $y_{d}$, and noise level $\sqrt{\bar\alpha}$ are inputs of model. The model estimates noise $\hat{\epsilon}$ to reconstruct $y_0$ from $y_{\bar\alpha}$.}
  \label{fig:model_figure}
\end{figure}

\subsection{Network architecture}
In this section, we introduce \nw, a conditional diffusion model for \asr. Our method directly generates waveform from waveform without feature extraction such as short-time Fourier transform spectrogram or phonetic posteriorgram. 
Figure \ref{fig:model_figure} illustrates the architecture of the \nw.
Based on the recent success of diffusion-based neural vocoders \textit{DiffWave} \cite{diffwave} and \textit{WaveGrad} \cite{wavegrad}, we engineer their architectures to fit \asr task.
To build a downsampled signal conditioned model, \nw model $\epsilon_{\theta}: \mathbb{R}^L\times\mathbb{R}^{L/r}\times\mathbb{R}\rightarrow \mathbb{R}^L$ takes the diffused signal $y_{\bar \alpha}\coloneqq\sqrt{\bar \alpha}\, y_0 +\sqrt{1 - \bar\alpha}\, \epsilon$, the downsampled signal $y_{d}$ and the noise level $\sqrt{\bar\alpha}$ as the inputs and its output $\hat\epsilon\coloneqq\epsilon_{\theta}(y_{\bar \alpha},y_d, \sqrt{\bar\alpha})$ estimates diffused noise $\epsilon$.
Similar to $\text{DiffWave}_{\text{BASE}}$ \cite{diffwave}, our model has $N=30$ residual layers with 64 channels. In each residual layer, bidirectional dilated convolution (\textit{Bi-DilConv}) with kernel size 3 and dilation cycle $[1,2,\dotsc,512]\times3$ is used.
Noise level embedding and local conditioner are added before and after the main Bi-DilConv to provide conditions of $\sqrt{\bar\alpha}$ and $y_{d}$.
After similar parts, there are two main modifications from the original DiffWave architecture: noise level embedding and local conditioner.

\noindent\textbf{Noise level embedding.}
DiffWave includes an integer timestep $t$ in the inputs to indicate a discrete noise level \cite{diffwave}.
On the other hand, inputs of WaveGrad contain a continuous noise level $\sqrt{\bar\alpha}$ to have a different number of iterations during inference \cite{wavegrad}. We implement the noise level embedding using 128-dimensional vector similar to sinusoidal positional encoding introduced in Transformer \cite{attention}:
\begin{align}
E(\sqrt{\bar\alpha}) = & \left[ \sin{\left(10^{-{[0:63]}\times\gamma} \times
         C\sqrt{\bar\alpha}\right)}, \right. \nonumber\\
         & \phantom{[}\left. \cos{\left(10^{-{[0:63]}\times\gamma} \times
         C\sqrt{\bar\alpha}\right)} \right].
\end{align}
We set a linear scale with $C=50000$ instead of 5000 introduced by Chen \etal \cite{wavegrad}, because the embedding with $C=5000$ cannot effectively distinguish among the low noise levels.
$\gamma$ in noise level embedding is set to 1/16.
The embedding is fed to the two shared fully connected (FC) layers and a layer-specific FC layer and then added as a bias term to the input of each residual layer. 

\noindent\textbf{Local conditioner.}
We modified the local conditioner to fit the audio upsampling task. After various experiments, we discovered that the receptive field for the conditional signal $y_{d}$ needs to be larger than the receptive field for the noisy input $y_{\bar\alpha}$. 
For each residual layer, we apply another Bi-DilConv with kernel size 3 for the conditioner and the same dilation cycle as the main Bi-DilConv to make the receptive field of $y_d$ nearly twice as large as that of $y_{\bar\alpha}$.
We hypothesize that having a large receptive field for the local conditions provides more useful information for upsampling.
While DiffWave upsampled the local condition with transposed convolution \cite{diffwave}, we adopt linear interpolation to build a single neural network architecture that could be utilized for different upscaling ratios.

\begin{figure}[t]
\setlength{\intextsep}{0pt}%
\begin{minipage}[t]{\linewidth}
\begin{algorithm}[H]
  \caption{\small Training.} \label{alg:training}
  \begin{algorithmic}[1]
    \algrenewcommand\algorithmicindent{1.0em}
    \Repeat
      \State $y_0 \sim q(y_0)$
      \State $y_d = \text{Subsampling}(\text{Filtering}(y_0),r)$
      \State $t \sim \mathrm{Uniform}(\{1,2, \dotsc, T\})$
      \State $\sqrt{\bar\alpha} \sim \mathrm{Uniform}(\sqrt{\bar\alpha_{t}}, \sqrt{\bar\alpha_{t-1}})$
      \State $\epsilon \sim \mathcal{N}(0,I)$
      \State Take gradient descent step on
      \Statex $\quad\; \nabla_\theta\! \log \left\lVert \epsilon - \epsilon_\theta\left(\!\sqrt{\bar\alpha}\, y_0 + \sqrt{1-\bar\alpha}\,\epsilon, y_{d}, \sqrt{\bar\alpha}\right) \right\rVert_1$
    \Until{converged}
  \end{algorithmic}
\end{algorithm}
\end{minipage}
\setlength{\intextsep}{10pt}%
\begin{minipage}[t]{\linewidth}
\begin{algorithm}[H]
  \caption{\small Sampling.}%
  \label{alg:sampling}
  \begin{algorithmic}[1]
    \algrenewcommand\algorithmicindent{1.0em}
    \State $y_T \sim \mathcal{N}(0, I)$
    \For{$t=T, T-1, \dotsc, 1$}
      \State $z \sim \mathcal{N}(0, I)$ if $t > 1$, else $z=0$%
      \State $\sigma_t = \sqrt{\frac{1-\bar\alpha_{t-1}}{1-\bar\alpha_t} \beta_t}$
      \State $y_{t-1} = \frac{1}{{\sqrt{\alpha_t}}}{\left(y_t - \frac{1-\alpha_t}{\sqrt{1-\bar\alpha_t}}\, \epsilon_\theta(y_t, y_{d}, \sqrt{\bar\alpha_t}) \right)}+ \sigma_t z$
    \EndFor
    \State \textbf{return} $\hat y= y_0$
  \end{algorithmic}
\end{algorithm}
\end{minipage}
\end{figure}
\subsection{Training objective}
We discovered that the $L_1$ norm 
scale of small timesteps and large timesteps differ by a factor of almost 10.
Substituting the $L_1$ norm to log-norm offers stable training results by scaling the losses from different timesteps.
Besides, \asr tasks need to substitute conditional factor from mel-spectrogram $x$ to downsampled signal $y_{d}$ as:
\begin{align}
    \mathbb{E}_{\bar\alpha,y_0,y_{d},\epsilon}\left[ \log \left\lVert  \epsilon  - 
    \epsilon_\theta\left(
    y_{\bar\alpha}
    , y_{d}, \sqrt{\bar\alpha}\right)\right\rVert_1 \right] . \label{eqn:nuwaveelbo}
\end{align}

Algorithm \ref{alg:training} illustrates the training procedure for \nw which is similar to the continuous noise level training suggested by Chen \etal \cite{wavegrad}.
The details of the filtering process are described in Section \ref{implementation detail}.
During sampling, we could utilize an untrained noise schedule with a different number of iterations by using continuous noise level training.
Thus, $T,\alpha_{1:T},\beta_{1:T}$ in Algorithm \ref{alg:training} and Algorithm \ref{alg:sampling} need not be identical.

\subsection{Noise schedule} \label{subsection:schedule}
We use linear noise schedule $\text{Linear}(1 \times 10^{-6}, 0.006, 1000)$ during training following the prior works on diffusion models \cite{ddpm, diffwave,wavegrad}.
We adjust the minimum and the maximum values from the schedule of WaveGrad $(\text{Linear}(1 \times 10^{-4}, 0.005, 1000))$ \cite{wavegrad} to train the model with a large range of noise levels.
Based on empirical results, $\sqrt{\bar\alpha_{T}}$ should be smaller than 0.5 to generate clean final output $y_0$.
During inference, the sampling time is proportional to a number of iterations $T$, thus we reduced it for fast sampling. 
We tested several manual schedules and found that 8 iterations with $\beta_{1:8} =[10^{-6},
        2\times 10^{-6} ,
        10^{-5} ,
        10^{-4} ,
        10^{-3} ,
        10^{-2} ,
        10^{-1} ,
        9\times 10^{-1} ]$ are suitable for our task.
Besides we did not report the test results with 1000 iterations as they are similar to that of 8 iterations.

\section{Experiments}
\subsection{Dataset}
We train the model on the VCTK dataset \cite{vctk} which contains 44 hours of recorded speech from 108 speakers.
Only the mic1 recordings of the VCTK dataset are used for experiments and recordings of \textit{p280} and \textit{p315} are excluded as conventional setup.
We divide \asr task into two parts: \textit{SingleSpeaker} and \textit{MultiSpeaker}. For the \textit{SingleSpeaker} task, we train the model on the first 223 recordings of the first speaker, who is labeled \textit{p225}, and test it on the last 8 recordings. For the \textit{MultiSpeaker} task, we train the model on the first 100 speakers and test it on the remaining 8 speakers.

\subsection{Training}
We train our model with the Adam \cite{adam} optimizer with the learning rate $3\times10^{-5}$.
The \textit{MultiSpeaker} model is trained with two NVIDIA A100 (40GB) GPUs, and the \textit{SingleSpeaker} model is trained with two NVIDIA V100 (32GB) GPUs.
We use the largest batch size that fits memory constraints, which is 36 and 24 for A100s and V100s.
We train our model for two upscaling ratios, $r=2 ,3$.
During training, we use the 0.682 seconds patches ($32768 - 32768 \textrm{ mod } r$ samples)
from the signals as the input, and during testing, we use the full signals.

\subsection{Evaluation}
To evaluate our results quantitatively, we measure the signal-to-noise ratio (SNR) and the log-spectral distance (LSD).
For a reference signal $y$ and an estimated signal $\hat y$, SNR is defined as $\text{SNR}(\hat y, y) =10 \log_{10} (\lVert y \rVert^2_2 /
     \lVert \hat y - y \rVert^2_2)$.
There are several works reported that SNR is not effective in the upsampling task because it could not measure high-frequency generation \cite{audiosuperres,bwegan}.
On the other hand, LSD could measure high-frequency generation as spectral distance.
Let us denote the log-spectral power magnitudes $Y$ and $\hat Y$ of signals $y$ and  $\hat y$, which is defined as $Y(\tau,k)=\log_{10} \left| S(y) \right|^2$ where $S$ is the short-time Fourier transform (STFT) with the Hanning window of size 2048 samples and hop length 512 samples, $\tau$ and $k$ are the time frame and the frequency bin of STFT spectrogram. The LSD is calculated as following:
\begin{align}
    \text{LSD}(\hat y, y) = \frac{1}{\mathcal{T}}\sum^{\mathcal{T}}_{\tau=1}\sqrt{\frac{1}{K} \sum^K_{k=1}\left(\hat Y(\tau,k)-  Y(\tau,k)\right)^2} .
    \label{eqn:lsd}
\end{align}

For qualitative measurement, we utilize the ABX test to determine whether the ground truth data and the generated data are distinguishable. In a test, human listeners are provided three audio samples called \textit{A}, \textit{B}, and \textit{X}, where \textit{A} is the reference signal, \textit{B} is the generated signal, and \textit{X} is the signal randomly selected between \textit{A} or \textit{B}. Listeners classify that \textit{X} is \textit{A} or \textit{B}, and we measure their accuracy. 
ABX test is hosted on the Amazon Mechanical Turk system, each of \textit{SingleSpeaker} and \textit{MultiSpeaker} task is examined by near 500 and 1500 cases.
We only tested it for the main upscaling ratio $r=2$.

\subsection{Baselines}
We compare our method with linear interpolation, U-Net \cite{unet} like model suggested by Kuleshov \etal \cite{audiosuperres}, and MU-GAN \cite{bwegan}.
To compare with our waveform-to-waveform model, we choose these models because U-Net is a basic waveform-to-waveform model and MU-GAN is a waveform-to-waveform GAN-based model with minimal auxiliary features in \asr.
We implemented and trained baseline models on same the 48kHz VCTK dataset.
While U-Net model was implemented as details in Kuleshov \etal \cite{audiosuperres}, we modified number of channels (down: $\text{min}(2^{b+2},32)$, up: $\text{min}(2^{12-b},128)$ for block index $b=1,2,...,8$) in MU-GAN for stable training.
We early-stopped training baseline models based on their LSD metrics
Each model's number of parameters is provided in Table \ref{tab:nop}, \nw only requires 5.4-21\% parameters than baselines.
Hyperparameters of baseline models might not be fully optimized for the 48kHz dataset. 
\begin{table}[th!]
  \caption{Comparison of model size. Our model has the smallest number of parameters as 5.4-21\% of other baselines.
  }
      \vspace{-0.5\baselineskip}
  \label{tab:nop}
  \centering
  \resizebox{0.9\linewidth}{!}{%
  \begin{tabular}{l c c c}
    \toprule
      & U-Net & MU-GAN & \nw (Ours)  \\
    \midrule
    \#parameter(M)$\downarrow$ & 56 & 14 & \textbf{3.0}  \\
    \bottomrule
  \end{tabular}}
\end{table}

\subsection{Implementation details} \label{implementation detail}
To downsample with in-phase, the filtering consists of STFT, fill zero to high-frequency elements, and inverse STFT (iSTFT).
STFT and iSTFT use the Hanning window of size 1024 and hop size 256.
The thresholds for high-frequency are the Nyquist frequency of the downsampled signals.
We cut out the leading and trailing signals 15dB lower than the max amplitude. 

\begin{figure*}[th!]
\setlength{\textfloatsep}{0pt}%

\newcommand{\incaptionimg}[3]{
  \begin{tikzpicture}[every node/.style={inner sep=0,outer sep=0}]
    \draw node[name=micrograph] {\includegraphics[width=\textwidth]{#1}}; 
    \draw  (micrograph.north west)  node[anchor=north west,yshift=-2.07cm,xshift=0.44cm,#3]{\textbf{\small{(#2)}}}; 
  \end{tikzpicture}
}
\newcommand{\incaptionlineimg}[4]{
  \begin{tikzpicture}[every node/.style={inner sep=0,outer sep=0}]
    \draw node[name=micrograph] {\includegraphics[width=\textwidth]{#1}}; 
    \draw[line width =0.5pt, red] (micrograph.south west)++(0.1*\textwidth,#4\textwidth)--++(0.895\textwidth,0); 
    \draw  (micrograph.north west)  node[anchor=north west,yshift=-2.07cm,xshift=0.44cm,#3]{\textbf{\small{(#2)}}}; 
  \end{tikzpicture}
}
\captionsetup[subfigure]{labelformat=empty}
    \setlength{\textfloatsep}{0pt}%
    \setlength{\intextsep}{0pt}
    \centering
    \begin{subfigure}[b]{0.2\linewidth}
    \subcaption{\quad Refernce Signal}
    \end{subfigure}%
    \begin{subfigure}[b]{0.2\linewidth}
    \subcaption{\quad Linear Interpolation}
    \end{subfigure}%
    \begin{subfigure}[b]{0.2\linewidth}
    \subcaption{\quad U-Net}
    \end{subfigure}%
    \begin{subfigure}[b]{0.2\linewidth}
    \subcaption{\quad MU-GAN}
    \end{subfigure}%
    \begin{subfigure}[b]{0.2\linewidth}
    \subcaption{\quad NU-Wave (ours)}
    \end{subfigure}\\
    \vspace{-0.80\baselineskip}
     \begin{subfigure}[b]{0.2\linewidth}
         \incaptionimg{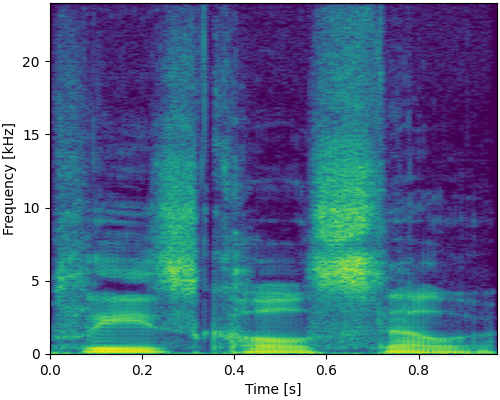}{a1}{white}
         \phantomsubcaption\ignorespaces\label{spec_a}
     \end{subfigure}%
     \begin{subfigure}[b]{0.2\linewidth}
         \incaptionlineimg{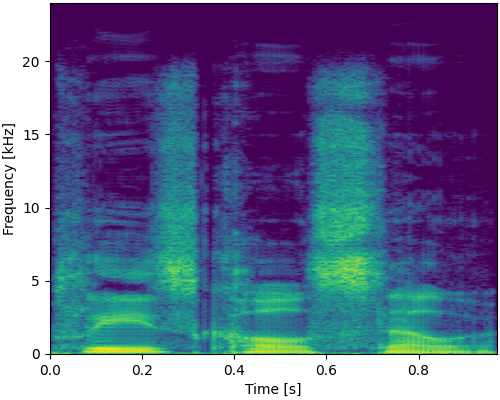}{a2}{white}{0.4425}
         \phantomsubcaption\ignorespaces\label{spec_b}
     \end{subfigure}%
     \begin{subfigure}[b]{0.2\linewidth}
         \incaptionlineimg{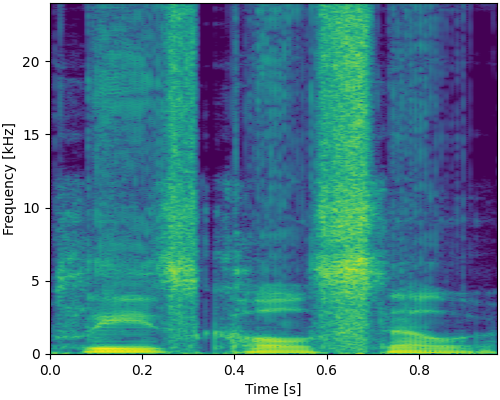}{a3}{white}{0.4425}
         \phantomsubcaption\ignorespaces\label{spec_c}
     \end{subfigure}%
     \begin{subfigure}[b]{0.2\linewidth}
         \incaptionlineimg{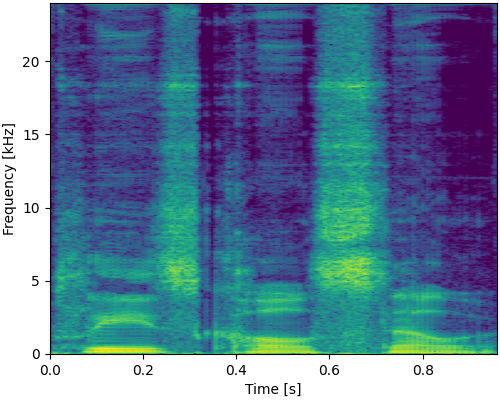}{a4}{white}{0.4425}
         \phantomsubcaption\ignorespaces\label{spec_d}
     \end{subfigure}%
     \begin{subfigure}[b]{0.2\linewidth}
         \incaptionlineimg{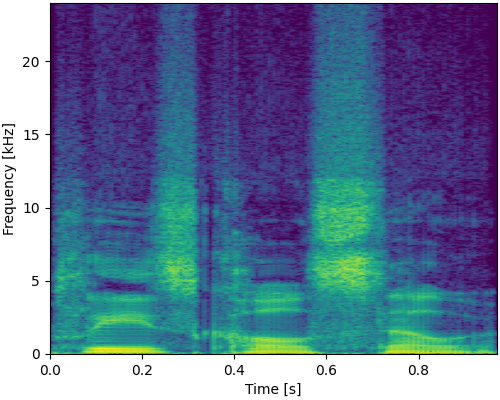}{a5}{white}{0.4425}
         \phantomsubcaption\ignorespaces\label{spec_e}
      \end{subfigure}\\
    \vspace{-1.2\baselineskip}
    \begin{subfigure}[b]{0.2\linewidth}
         \incaptionimg{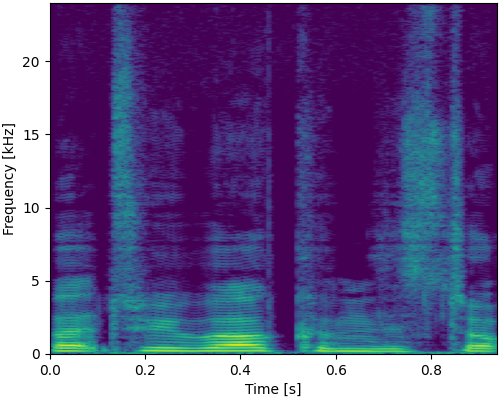}{b1}{white}
         \phantomsubcaption\ignorespaces\label{spec_f}
     \end{subfigure}%
     \begin{subfigure}[b]{0.2\linewidth}
         \incaptionlineimg{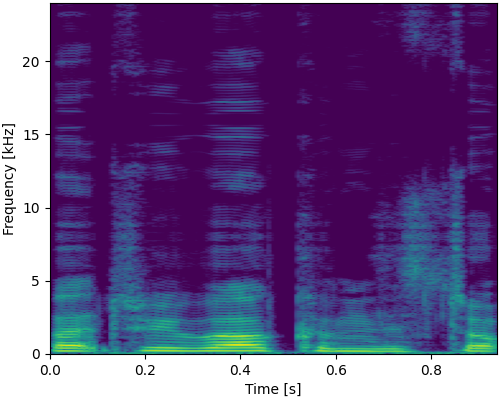}{b2}{white}{0.326}
         \phantomsubcaption\ignorespaces\label{spec_g}
     \end{subfigure}%
     \begin{subfigure}[b]{0.2\linewidth}
         \incaptionlineimg{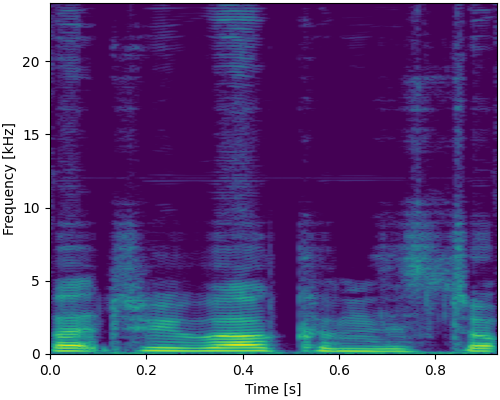}{b3}{white}{0.326}
         \phantomsubcaption\ignorespaces\label{spec_h}
     \end{subfigure}%
     \begin{subfigure}[b]{0.2\linewidth}
         \incaptionlineimg{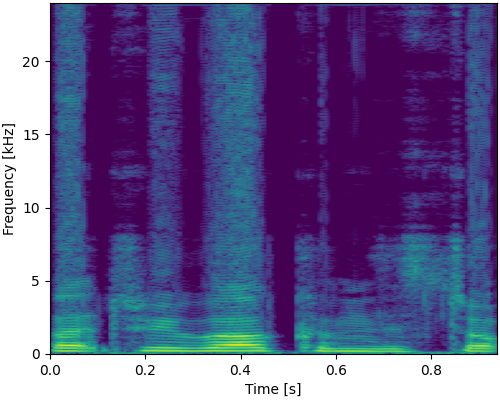}{b4}{white}{0.326}
         \phantomsubcaption\ignorespaces\label{spec_i}
     \end{subfigure}%
     \begin{subfigure}[b]{0.2\linewidth}
         \incaptionlineimg{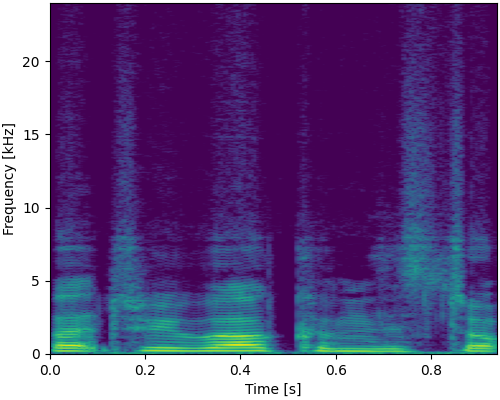}{b5}{white}{0.326}
         \phantomsubcaption\ignorespaces\label{spec_j}
     \end{subfigure}%
    \vspace{-1.8\baselineskip}
    \caption{Spectrograms of reference and upsampled speeches. Red lines indicate the Nyquist frequency of downsampled signals.\\\textbf{\textup{(a1-a5)}} are samples of $r=2$, MultiSpeaker (p360\_001) and \textbf{\textup{(b1-b5)}} are samples of $r=3$, SingleSpeaker (p225\_359). More samples are available at \url{https://mindslab-ai.github.io/nuwave}
    }
    \label{fig:spec}
\end{figure*}

\section{Results}
As illustrated in Figure \ref{fig:spec},
our model can naturally reconstruct high-frequency elements of sibilants and fricatives,
while baselines tend to generate that resemble low-frequency elements by flipping along with downsampled signal's Nyquist frequency.

Table \ref{tab:quant} shows the quantitative evaluation of our model and the baselines. 
For all metrics, \nw outperforms the baselines.
Our model improves SNR value by 0.18-0.9 dB from the best performing baseline, MU-GAN. The LSD result achieves almost half ($r=2$: 43.8-45.0\%, $r=3$: 55.6-57.3\%) of linear interpolation's LSD, while the best baseline model (MU-GAN, $r=2$, \textit{SingleSpeaker}) only achieves 62.2\%.
We tested our model five times and observed that the standard deviations of the metrics are remarkably small, in the factor of $10^{-5}$.
\begin{table}[th]
  \caption{Results of evaluation metrics. Upscaling ratio ($r$) is indicated as $\times 2, \times 3$.
  Our model outperforms other baselines for both SNR and LSD.
  }
    \vspace{-0.5\baselineskip}

  \label{tab:quant}
  \centering
  \resizebox{0.9\linewidth}{!}{%
  \begin{tabular}{l@{}l cccc}
  \toprule
  \multicolumn{2}{l}{\multirow{2}{*}{\textbf{Model}}} & \multicolumn{2}{c}{SingleSpeaker} &\multicolumn{2}{c}{MultiSpeaker}\\
  \cmidrule{3-6}
        & & SNR $ \uparrow $ & LSD $ \downarrow $ & SNR $ \uparrow $ & LSD $ \downarrow $ \\
  \midrule
    Linear&$\times 2$ 
                        & $\phantom{0}9.69$ & $1.80\phantom{0}$  & $11.1\phantom{0}$ & $1.93\phantom{0}$   
    \\
    U-Net&$\times 2$ 
                        &$10.3\phantom{0}$ & $1.57\phantom{0}$ & $\phantom{0}9.86$ & $1.47\phantom{0}$   
    \\
    MU-GAN&$\times 2$ 
                        & $10.5\phantom{0}$ & $1.12\phantom{0}$ & $12.3\phantom{0}$ & $1.22\phantom{0}$
    \\
    \nw&$\times 2$       
                        & $\mathbf{11.1\phantom{0}}$ & $\mathbf{0.810}$ & $\mathbf{13.2}\phantom{0}$ & $\mathbf{0.845}$
    \\
    \midrule
    Linear&$\times 3$ 
                        & $8.04$ & $1.72\phantom{0}$  & $\phantom{0}8.71$ & $1.74\phantom{0}$  
    \\
    U-Net&$\times 3$ 
                        & $8.81$  & $1.64\phantom{0}$
                        & $10.7\phantom{0}$ & $1.41\phantom{0}$   
    \\
    MU-GAN&$\times 3$ 
                        & $9.44$ & $1.37\phantom{0}$  & $11.7\phantom{0}$ & $1.53\phantom{0}$   
    \\
    \nw&$\times 3$       
                        & $\mathbf{9.62}$ & $\mathbf{0.957} $ & $\mathbf{12.0}\phantom{0}$ & $\mathbf{0.997}$
    \\
  \bottomrule
  \end{tabular}}
\end{table}
\begin{table}[th]
  \caption{The accuracy and the confidence interval of the ABX test. \nw shows the lowest accuracy (51.2-52.1\%) indicating that its outputs are indistinguishable from the reference.}
      \vspace{-0.5\baselineskip}
  \label{tab:abx}
  \centering
    \resizebox{0.9\linewidth}{!}{%
  \begin{tabular}{ l @{}l c c}

    \toprule
    \multicolumn{2}{l}{\textbf{Model}} &   SingleSpeaker & MultiSpeaker
    \\
    \midrule
    Linear&$\times 2$ 
                        & $55.5\pm0.04$\% & $52.3\pm0.03$\%~~~
    \\
    U-Net&$\times 2$ 
                        & $52.9\pm0.04$\% & $52.5\pm0.03$\%~~~
    \\
    MU-GAN&$\times 2$ 
                        & $\mathbf{52.1}\pm0.04$\% & $\mathbf{51.3}\pm0.03$\%~~~
    \\
    \nw&$\times 2$       
                        & $\mathbf{52.1}\pm0.04$\% & $\mathbf{51.2}\pm0.03$\%~~~
    \\
    \bottomrule
  \end{tabular}}
\end{table}

Table \ref{tab:abx} shows the accuracy of the ABX test.
Since our model achieves the lowest accuracy (52.1\%, 51.2\%) close to 50\%,
we can confidently claim that \nw's outputs are almost indistinguishable from the reference signals.
While it is notable to mention that the confidence interval of our model and MU-GAN is mostly overlapped, the difference in accuracy was not statistically significant.

\section{Discussion}
In this paper, we applied a conditional diffusion model for \asr. To the best of our knowledge, \nw is the first \asr model that produces high-resolution 48kHz samples from 16kHz or 24kHz speech, and the first model that successfully utilized the diffusion model for the audio upsampling task. 

\nw outperforms other models in quantitative and qualitative metrics with fewer parameters.
While other baseline models obtained the LSD similar to that of linear interpolation, our model achieved almost half the LSD of linear interpolation. This indicates that \nw can generate more natural high-frequency elements than other models. 
The ABX test results show that our samples are almost indistinguishable from reference signals.
However, the differences between the models were not statistically significant, since the downsampled signals already contain the information up to 12kHz which has the dominant energy within the audible frequency band.
Since our model outperforms baselines for both \textit{SingleSpeaker} and \textit{MultiSpeaker}, we can claim that our model generates high-quality upsampled speech for both the seen and the unseen speakers. 

While our model generates natural sibilants and fricatives, it cannot generate harmonics of vowels as well.
In addition, our samples contain slight high-frequency noise.
In further studies, we can utilize additional features, such as pitch, speaker embedding, and phonetic posteriorgram \cite{asr4sr,spkppgbwe}.
We can also apply more sophisticated diffusion models, such as DDIM \cite{ddim}, CAS \cite{cas}, or VP SDE \cite{sde}, to reduce upsampling noise.

\section{Acknowledgements}
The authors would like to thank Sang Hoon Woo and teammates of MINDs Lab, 
Jinwoo Kim, Hyeonuk Nam from KAIST, and Seung-won Park from SNU for valuable discussions.

\bibliographystyle{IEEEtran}

\bibliography{main}

\end{document}